\documentclass{article}
\usepackage[a4paper, total={6in, 8in}]{geometry}
\usepackage{setspace} 
\usepackage{mathrsfs}
\usepackage{bbm}

\usepackage{dutchcal}
\usepackage{verbatim}
\usepackage{amsmath}
\usepackage{amsthm}
\usepackage{amssymb}
\usepackage{mathtools}
\usepackage{bm}
\usepackage{multirow}
\usepackage{graphicx}
\usepackage{float}
\usepackage{comment}
\usepackage{amsfonts}
\usepackage{color}
\usepackage[dvipsnames]{xcolor}
\usepackage{xr-hyper}
\usepackage[colorlinks=true,linkcolor=blue,citecolor=blue,pdfborder={0 0 0}]{hyperref}
\usepackage[backend=bibtex,style=ieee,maxbibnames=99,natbib=true,citetracker=true]{biblatex}
\usepackage{cleveref}
\usepackage[affil-it]{authblk}

\makeatletter
\newcommand{\refcheckize}[1]{%
  \expandafter\let\csname @@\string#1\endcsname#1%
  \expandafter\DeclareRobustCommand\csname relax\string#1\endcsname[1]{%
    \csname @@\string#1\endcsname{##1}\wrtusdrf{##1}}%
  \expandafter\let\expandafter#1\csname relax\string#1\endcsname
}
\makeatother

\usepackage{graphicx}
\graphicspath{ {Images/20211221/} }
\usepackage{caption}
\usepackage{subcaption}

\numberwithin{equation}{section}

\theoremstyle{plain}
\newtheorem{Assump}{Assumption}[section]

\newtheorem{Prop}{Proposition}[section]

\theoremstyle{remark}

\newtheorem{Remark}{Remark}[section]

\DeclareMathOperator{\Var}{Var}
\DeclareMathOperator{\Cov}{Cov}

\newcommand{\ca}{\mathcal{a}}

\newcommand{\tb}{\tilde{b}}

\newcommand{\hA}{\hat{A}}
\newcommand{\hC}{\hat{C}}
\newcommand{\hF}{\hat{F}}
\newcommand{\htF}{\hat{\tilde{F}}}
\newcommand{\tM}{\tilde{M}}
\newcommand{\hS}{\hat{S}}
\newcommand{\cS}{\mathcal{S}}
\newcommand{\hcS}{\hat{\mathcal{S}}}

\newcommand{\htU}{\hat{\tilde{U}}}
\newcommand{\hU}{\hat{U}}

\newcommand{\bx}{\bm{x}}
\newcommand{\bbZ}{\mathbbm{Z}}
\newcommand{\ind}{\mathbbm{1}}
\newcommand*\diff{\mathop{}\!\mathrm{d}}
\newcommand{\Exp}{\mathbb{E}}
\newcommand{\Prob}{\mathbb{P}}

\newcommand{\floor}[1]{\lfloor{#1}\rfloor}

\title{POT-flavored estimator of Pickands dependence function}
\date{}
\author{Nan Zou\thanks{Email: nan.zou@mq.edu.au}}
\affil{School of Mathematical and Physical Sciences, Macquarie University, Sydney}
\begin{document}
\maketitle
\begin{abstract}
    
This work proposes an estimator with both Peak-Over-Threshold and Block-Maxima flavors, uses it to estimate the Pickands dependence function of bivariate time series, and illustrates how it brings down the asymptotic bias and the overall mean squared error.

\end{abstract}

\section{Introduction}

Extreme value statistics is witnessing an intensive horse racing \cite{bucher2021horse} between two fundamental methods: the Block Maxima (BM) method and the Peak-Over-Threshold (POT) method. Intuitively, The BM method partitions the observations into blocks and view the max of each block to be extreme, while the POT method sets a threshold and considers the observations above this threshold to be extreme. The BM method and the POT method are connected but not identical to each other.  

In addition to asking which of the BM and POT methods prevails over the other, one may also wonder if these two methods could be mixed to obtain an even better performance. This manuscript aims to propose a estimator that have the flavor of both BM and POT methods in the bivariate time series setting. 
Consider strictly stationary bivariate time series $X_{t}=\{X_{t,1}, X_{t,2}\}_{t\in \bbZ}$ with continuous univariate stationary margins. First we refer to the BM method. For $j=1,2$, let 
\begin{equation}\label{eq:blockMax}
    M_{m,1,j} = \max \{X_{t,j}: t=1,\dots,m\}
\end{equation}
be the coordinate-wise block maxima, $M_{m,1}=(M_{m,1,1},M_{m,1,2})$ be the block maxima vector, and $C_{m}$ be the copula of $M_{m,1}$. Assume
there exists an extreme copula $C_{\infty}$ such that for all $(u,v)\in [0,1]^{2}$,  
\begin{equation}\label{eq:extcopula}
    \lim_{m\to \infty} C_{m}(u,v) = C_{\infty}(u,v).
\end{equation}
Indeed, under \eqref{eq:extcopula}, $C_{\infty}$ can be represented with some function $A_{\infty}:[0,1]\to [0,1]$; for $(u,v)\in [0,1]^{2}$,
\begin{equation}\label{eq:copula_in_pickands}
C_{\infty}(u,v) = \exp\Bigg\{\log(uv)A_{\infty}\bigg(\frac{\log v}{\log (uv)}\bigg)\Bigg\}. 
\end{equation}
Hence, the inference of the extreme copula $C_{\infty}$ boils down to the inference of $A_{\infty}$, which is called the Pickands dependence function of $C_{\infty}$. The estimation of the Pickand function $A_{\infty}$ has drawn a considerable attention in the literature, e.g., \cite{Pic81,caperaa1997nonparametric,hall2000distribution,genest2009rank,gudendorf2012nonparametric,bucher2011new,berghaus2013minimum,peng2013weighted,cormier2014using,escobar2018local}; for an overview, see \cite{segers2012nonparametric,vettori2018comparison}. In particular, 
\cite{naveau2009modelling,guillou2014madogram,fonseca2015generalized,marcon2017multivariate} develop fast-to-compute, easy-to-interpret, madogram-type estimators. 

Most of the literature above postulate that $C_{m}= C_{\infty}$ instead of \eqref{eq:extcopula} and as a result does not include the bias, $C_{m}-C_{\infty}$, in their asymptotic analyses. To cope with this bias term, one way is to push the block size $m$ to infinity; alternatively, one could, as in the POT method, potentially consider pushing the threshold to infinity. In this work, instead of pushing the threshold to infinity and only considering the observations above the threshold as in the POT method, we propose to put more ``weight'' on the larger observations and develop a POT-flavored, madogram-type estimator for the Pickand dependent function $A_{\infty}$ in \eqref{eq:copula_in_pickands} in the bivariate time series setting. With asymptotic analyses and simulation on the bias and the variance under \eqref{eq:extcopula}, we find that in some scenarios assigning more weight on larger observations, although increases the variance, can reduce the bias and can moreover bring down the overall mean squared error (MSE).

The remaining parts of this paper proceed as follows. \Cref{sec:Pick} details the construction of this POT-flavored Pickands dependence function estimator. \Cref{sec:asymp} analyzes the asymptotic property of this estimator for the Pickands dependence function. \Cref{sec:copula} discusses the choice of the copula estimators. \Cref{sec:simul} presents the simulation results. The Appendix includes all the proofs.

\section{POT-flavored Pickands Dependence Function Estimator}\label{sec:Pick}
Let 
\begin{equation}\label{eq:mado_in_copula0}
    \cS(t) = 1-\int_{0}^{1}C_{\infty}(y^{1-t}, y^{t})\diff y. 
\end{equation}
When designing madogram-type estimators for the Pickands dependence function $A_{\infty}$ in \eqref{eq:copula_in_pickands}, \cite{naveau2009modelling,guillou2014madogram,fonseca2015generalized,marcon2017multivariate} leverage the fact that \eqref{eq:mado_in_copula0} gives
\[
    A_{\infty}(t) = \frac{1}{c}\bigg(\frac{1}{1-\cS(t)}-1\bigg).
\]
To assign more weights to those $C_{\infty}(u,v)$ evaluated at larger values of $u,v \in [0,1]^{2}$ in \eqref{eq:mado_in_copula0}, for $t\in [0,1]$ and $c>0$, we let $\cS(t,c)$, a generalization of $\cS(t)$, be defined by
\begin{equation}\label{eq:mado_in_copula}
    \cS(t,c) = 1-\int_{0}^{1}C_{\infty}(y^{c(1-t)}, y^{ct})\diff y. 
\end{equation}
Plugging \eqref{eq:copula_in_pickands} into \eqref{eq:mado_in_copula} gives that for all $c>0$,
\begin{equation}\label{eq:pickands_in_mado}
A_{\infty}(t) = \frac{1}{c}\bigg(\frac{1}{1-\cS(t,c)}-1\bigg).
\end{equation}
As an empirical counterpart of \eqref{eq:mado_in_copula} and \eqref{eq:pickands_in_mado}, let $\hcS(t,c)$, the estimator for $\cS(t,c)$, be defined by 
\begin{equation}\label{eq:hat_madogram_in_hat_copula}
\hcS(t,c) = 1-\int_{0}^{1}\hC_{\infty}(y^{c(1-t)}, y^{ct})\diff y,
\end{equation}
and let $\hA_{\infty}(t)$, the estimator for $A_{\infty}(t)$, be defined by
\[
\hA_{\infty}(t) = \frac{1}{c}\bigg(\frac{1}{1-\hcS(t,c)}-1\bigg).
\]
\begin{Remark}\label{rem:mado}
Notice that $\cS(t,c)$ connects closely to the
mean of the maximum of variables and the mean of the absolute difference of variables. 
Indeed, for $j=1,2$, let 
\begin{equation}\label{eq:marginal}
\begin{aligned}
F_{m,j}(x) &= \Prob(M_{m,1,j}\leq x), & U_{m,1,j} &= F_{m,j}(M_{m,1,j}),
\end{aligned}
\end{equation}
where $M_{m,1,j}$ is defined in \eqref{eq:blockMax}. Then, by \eqref{eq:extcopula}, \eqref{eq:mado_in_copula}, and the tail sum formula for expectation,
\begin{align*}
\cS(t,c) 
=\lim_{m\to \infty} \Exp \left [\max \left(U_{m,1,1}^{\frac{1}{c(1-t)}},U_{m,1,2}^{\frac{1}{ct}}\right)\right]=\lim_{m\to \infty} \frac{1}{2}\Exp \left[ U_{m,1,1}^{\frac{1}{c(1-t)}}+U_{m,1,2}^{\frac{1}{ct}}+\left|U_{m,1,1}^{\frac{1}{c(1-t)}}-U_{m,1,2}^{\frac{1}{ct}}\right|\right].
\end{align*}
\end{Remark}

\begin{Remark}\label{rem:POT}
By a change of variables, \eqref{eq:hat_madogram_in_hat_copula} results in 
\[
\hcS(t,c) =  1-\int_{0}^{1}\hC_{\infty}(y^{(1-t)}, y^{t})\frac{1}{c}y^{1/c-1}\diff y.
\]
Hence, when $c$ gets smaller, the integral in the expression of $\hcS(t,c)$ put more weights on those $\hC_{\infty}(y^{(1-t)}, y^{t})$ with larger value of $y$. 
Hence, when $c$ gets smaller, the values of $\hC_{\infty}(u,v)$ at larger values of $u$ and $v$ will take more weight in the construction of $\hA_{\infty}$. As a result, $\hA_{\infty}$ has some flavor of the POT. 
\end{Remark}

\section{Asymptotic Properties}\label{sec:asymp}
By the Continuous Mapping Theorem, the definition of equicontinuity, a Taylor expansion, and the Slutsky's Theorem, for fixed $c>0$, $\hcS(\cdot,c)$ and $\hA_{\infty}(\cdot)$ will be consistent and asymptotically Gaussian if the copula estimator $\hC_{\infty}(\cdot)$ is consistent and asymptotically Gaussian. The asymptotic bias and variance of $\hA_{\infty}$ depend on the specific choice of $\hC_{\infty}$. For simplicity, we choose $\hC_{\infty}$ to be the disjoint-block copula estimator  $\hC_{m}^{\circ}$ in, e.g., \cite{Buc14}. More specifically, recall $F_{m,j}, j=1,2$, defined in \eqref{eq:marginal}. Let
\begin{equation}\label{eq:C_circ}
\begin{aligned}
    \tM_{m,i,j} &= \max \{X_{t,j}:t\in [(i-1)m+1,im]\cap \bbZ\}  &
    U_{m,i,j} &=F_{m,j}(\tM_{m,i,j})\\ \tb&=\floor{n/m} &
    \hC_{m}^{\circ}(u,v)&= \frac{1}{\tb}\sum_{i=1}^{\tb}\ind (U_{m,i,1}\leq u, U_{m,i,2}\leq v).
\end{aligned}
\end{equation}

\begin{Assump}\label{assump:high_order}
We assume there exists a positive function $\ca(\cdot)$ with $\lim_{m\to\infty} \ca(m)=0$ and a non-null function $S$ on $[0,1]^{2}$ such that
\begin{align} \label{eq:secor}
\lim_{m\to\infty} \frac{C_{m}(u,v)-C_\infty(u,v)}{\ca(m)}  = S(u,v)
\end{align}
uniformly in $(u,v)\in[0,1]^2$. Subsequently, we assume that $\ca(\cdot)$ is regularly varying of order $\rho< 0$, that is, $\lim_{m\to\infty} \ca(mx)/\ca(m)= x^\rho$ for all $x>0$. 
\end{Assump}

\begin{Assump}\label{assump:AlphaMixing}
Assume $\{X_{t}\}_{t\in \bbZ}$ to be $\alpha$-mixing with coefficient $\alpha(k), k=1,2,\dots$. Further, assume that, as $n\to\infty$, there exists a positive integer sequence $\ell=\ell_{n}$ such that 
\begin{align*}
    &(i) \ m\to\infty, \ m=o(n) &
    &(ii) \ \ell\to\infty, \ \ell=o(m)\\
    &(iii) \ (n/m)\alpha(\ell) = o(1), \ (m/\ell)\alpha(\ell)=o(1)&
    &(iv) \ \alpha(k)= O(k^{-(1+\varrho)}) \ \text{for some} \ \varrho>0.
\end{align*}
\end{Assump}

\begin{Prop}\label{prop:BiasVar}
Under \Cref{assump:high_order}, \Cref{assump:AlphaMixing}, and \eqref{eq:C_circ},
\begin{align*}
    &(i) \ \left|\Exp\big[\hA_{\infty}(t)-A_{\infty}(t)\big]\right|=\ca(m)S\Big(e^{-(1-t)},e^{-t}\Big)e^{A_{\infty}(t)}\Gamma(2-\rho)\Bigg(\frac{cA_{\infty}(t)+1}{c}\Bigg)^{\rho}+o(\ca(m));\\
     &(ii) \ \Var\big(\hA_{\infty}(t)\big)=(m/n)\frac{\big(cA_{\infty}(t)+1\big)^{2}A_{\infty}(t)}{c\big(cA_{\infty}(t)+2\big)}+o(m/n),
\end{align*}
where $\Gamma(\cdot) $ is the Gamma function.
\end{Prop}
\begin{Remark}
Since $\rho<0$, by analyzing the derivatives, the dominating terms of $\left|\Exp\big[\hA_{\infty}(t)-A_{\infty}(t)\big]\right|$ and $\Var\big(\hA_{\infty}(t)\big)$ turn out to be an increasing and a decreasing function, respectively, with respect to the constant $c$. Recall that by \Cref{rem:POT}, a smaller constant $c$ leads to a larger weights for higher values, or intuitively, a higher ``threshold''. Hence, \Cref{prop:BiasVar} indicates that when this ``threshold'' gets higher, the absolute value of bias of the estimator will be smaller while the variance will become larger.  
\end{Remark}
\begin{Remark}
In light of the POT method, we can set $c=c_{n}\to 0$, namely, we can let the ``threshold'' goes to infinity as $n\to \infty$. In this case, both the orders of $\left|\Exp\big[\hA_{\infty}(t)-A_{\infty}(t)\big]\right|$ and $\Var\big(\hA_{\infty}(t)\big)$ depend on the ratio $m/c$. Specifically, the absolute value of the bias will have an order of $\ca(m/c)$ and the variance will have an order of $(m/c)/n$. 
\end{Remark}

\section{Choice of Copula Estimator}\label{sec:copula}
In practice, we can substitute the disjoint-block (denoted by D) estimator of \cite{Buc14} and the overlapping-block (denoted by O) estimator of \cite{zou2021multiple} for $\hC_{\infty}$ in \eqref{eq:hat_madogram_in_hat_copula}. Specifically, recall $\tb$ and $\tM_{m,i,j}$ in \eqref{eq:C_circ}. Let $b=n-m+1$, $M_{m,i,j} = \max \{X_{t,j}:t\in [i,i+m-1]\cap \bbZ\}$, and $\hC_{m}^{D}$ and $\hC_{m}^{O}$ be the disjoint-block and overlapping-block estimator, respectively, defined by:
\begin{align*}
\htF_{m,j}(x)&=\frac{1}{\tb}\sum_{i=1}^{\tb}\ind(\tM_{m,i,j}\leq x)&\hF_{m,j}(x)&=\frac{1}{b}\sum_{i=1}^{b}\ind(M_{m,i,j}\leq x)
\\
\htU_{m,i,j}&=\htF_{m,j}(\tM_{m,i,j}) & \hU_{m,i,j}&=\hF_{m,j}(M_{m,i,j})\\
\hC_{m}^{D}(u,v)&= \frac{1}{\tb}\sum_{i=1}^{\tb}\ind (\htU_{m,i,1}\leq u, \htU_{m,i,2}\leq v) & \hC_{m}^{O}(u,v)&= \frac{1}{b}\sum_{i=1}^{b}\ind (\hU_{m,i,1}\leq u, \hU_{m,i,2}\leq v).
\end{align*}
By plugging $\hC_{m}^{D}$ and $\hC_{m}^{O}$ back to \eqref{eq:hat_madogram_in_hat_copula}, for $\Psi=D,O$, we can define estimators for the Pickands dependence function by
\[
\hcS_{m}^{\Psi}(t,c) = 1-\int_{0}^{1}\hC_{m}^{\Psi}(y^{c(1-t)}, y^{ct})\diff y.
\]

\begin{Remark}
Similar to \Cref{rem:mado}, 
\begin{equation}\label{eq:hat_mado_simu}
\begin{aligned}
\hcS_{m}^{D}(t,c)&=\frac{1}{\tb}\sum_{i=1}^{\tb}\max (\htU_{m,i,1}^{\frac{1}{c(1-t)}}, \htU_{m,i,2}^{\frac{1}{ct}})&
\hcS_{m}^{O}(t,c)&=\frac{1}{b}\sum_{i=1}^{b}\max (\hU_{m,i,1}^{\frac{1}{c(1-t)}}, \hU_{m,i,2}^{\frac{1}{ct}}).
\end{aligned}
\end{equation}

\end{Remark}

\section{Simulation}\label{sec:simul}
\subsection{Data Generating Process}
Let $n = 1000$ be the sample size. We consider the moving maximum processes in the setup section of Chapter 5 of \cite{Buc14}. In particular, we let 
\begin{align*}
    X_{t,1} = \max (W_{t,1}^{1/a}, W_{t-1,1}^{1/(1-a)}), X_{t,2} = \max (W_{t,2}^{1/b}, W_{t-1,2}^{1/(1-b)}), 
\end{align*}
where we set $a=0.25$, $b=0.5$, and let $\{W_{t,1}, W_{t,2}\}_{t\in \bbZ}$ be a bivariate iid sequence with uniform marginal distributions on $[0,1]$ and a joint cumulative distribution function $D$ specified below. 

\subsubsection{Outer-power Transformation of Clayton Copula}\label{sec:OuterPowerClayton}
The outer-power transformation of a Clayton copula is defined, for $(u,v)\in [0,1]^{2}$, by
\[
D
(u,v) = \left[1+\{(u^{-\theta}-1)^{\beta}+(v^{-\theta}-1)^{\beta}\}^{1/\beta}\right]^{-1/\theta},\]
where we set $\theta= 1$ and $\beta =\log(2)/\log(2-0.25)$.
\subsubsection{$t$-Copula}\label{sec:tCopula}
The $t$-copula is defined, for $(u,v)\in [0,1]^{2}$, as
\begin{equation*}
{\small D(u,v
) 
=\int_{-\infty}^{t_{\nu}^{-1}(u)} \int_{-\infty}^{t_{\nu}^{-1}(v)}\frac{\Gamma\Big(\frac{\nu+2}{2}\Big)}{\Gamma\big(\frac{\nu}{2}\big)\pi\nu |P|^{1/2}}\bigg(1+\frac{\bx'P^{-1}\bx}{\nu}\bigg)^{-\frac{\nu+2}{2}}\diff x_2 \diff x_1, }
\end{equation*}
where $\bx=(x_{1},x_{2})'$, $P$ is a $2\times 2$ correlation matrix with off-diagonal element $\theta$, and $t_{\nu}$ is the cumulative distribution function of a standard univariate $t$-distribution with degrees of freedom $\nu$. We set $\nu=4$ and $\theta=0.494217$ so that the coefficient of upper tail dependence of $D$ matches the coefficient in the outer-power transformation of Clayton Copula in \Cref{sec:OuterPowerClayton}.

\subsubsection{Gaussian Copula}

The Gaussian copula is defined, for $(u,v)\in [0,1]^{2}$, as
\begin{equation*}
{\small D(u,v
) 
=\int_{-\infty}^{\Phi^{-1}(u)} \int_{-\infty}^{\Phi^{-1}(v)}\frac{1}{2\pi |P|^{1/2}}\exp\bigg(-\frac{\bx'P^{-1}\bx}{2}\bigg)\diff x_2 \diff x_1, }
\end{equation*}
where $\bx=(x_{1},x_{2})'$, $P$ is a $2\times 2$ correlation matrix with off-diagonal element $\theta$, and $\Phi$ is the cumulative distribution function of a standard univariate normal distribution. We set $\theta=0.5$ to match the coefficient in the $t$-copula in \Cref{sec:tCopula}.

\subsection{Algorithm}
Choose block size $m = 1,2,\dots, 30$. We estimate the Pickands Dependence Function $A_{\infty}(t)$ with the additive boundary corrected estimator $\check{A}_{m}^{\Psi}(t)$. Specifically, for $\Psi=D, O,$ define
\[
\check{A}_{m}^{\Psi}(t)=\hA_{m}^{\Psi}(t)-(1-t)\{\hA_{m}^{\Psi}(0)-1\}-t\{\hA_{m}^{\Psi}(1)-1\},
\]
where 
\begin{equation*}
\hA_{m}^{\Psi}(t) = \frac{1}{c}\bigg(\frac{1}{1-\hcS_{m}^{\Psi}(t,c)}-1\bigg),
\end{equation*}
where $\hS_{m}^{\Psi}(t,c)$, $\Psi=D, O,$ are generated by \eqref{eq:hat_mado_simu}.

\subsection{Result}
\begin{figure}[H]
\centering
\begin{subfigure}{.3\textwidth}
  \centering
  \includegraphics[width=1.0\linewidth]{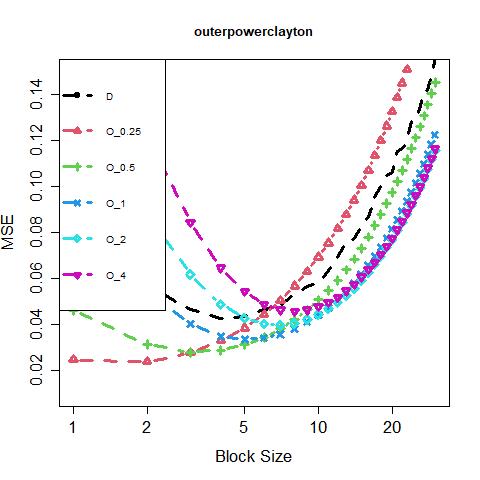}
\end{subfigure}%
\begin{subfigure}{.3\textwidth}
  \centering
  \includegraphics[width=1.0\linewidth]{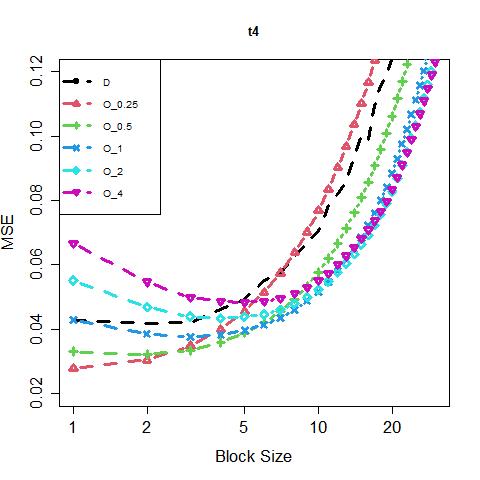}
\end{subfigure}%
\begin{subfigure}{.3\textwidth}
  \centering
  \includegraphics[width=1.0\linewidth]{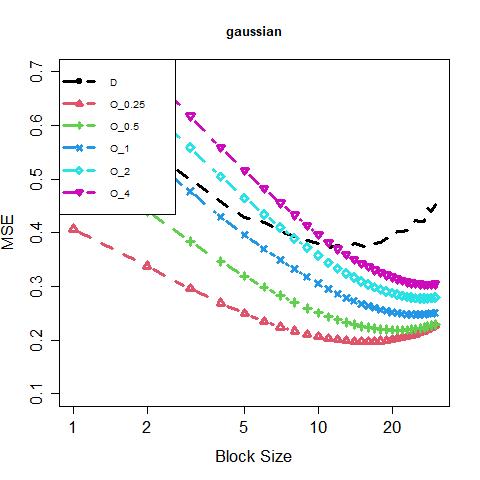}
\end{subfigure}
\caption{$MSE^{(sum)}$ of estimators in \Cref{tab:est} under outer-power Clayton, $t_{4}$, and Gaussian copulas}
\label{fig:mse}
\end{figure}

\begin{figure}[H]
\centering
\begin{subfigure}{.3\textwidth}
  \centering
  \includegraphics[width=1.0\linewidth]{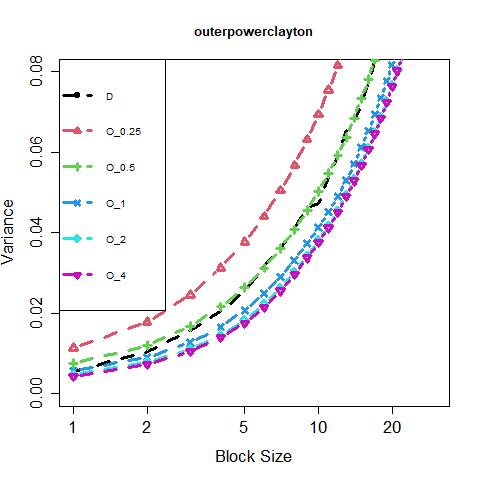}
\end{subfigure}%
\begin{subfigure}{.3\textwidth}
  \centering
  \includegraphics[width=1.0\linewidth]{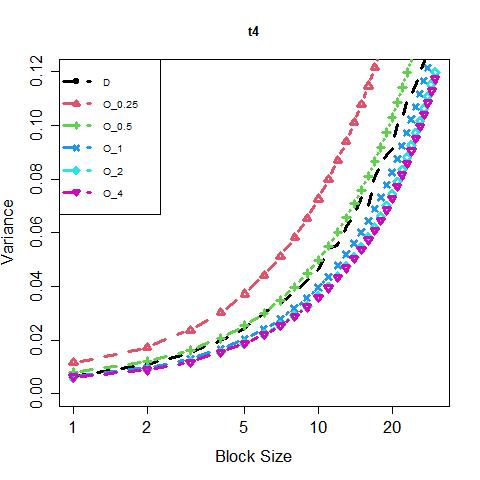}
\end{subfigure}%
\begin{subfigure}{.3\textwidth}
  \centering
  \includegraphics[width=1.0\linewidth]{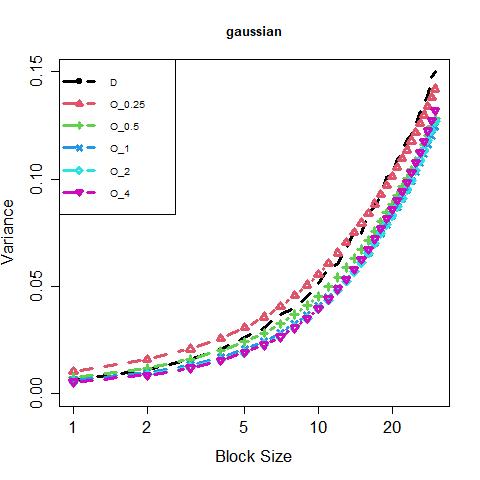}
\end{subfigure}
\caption{$Var^{(sum)}$ of estimators in \Cref{tab:est} under outer-power Clayton, $t_{4}$, and Gaussian copulas}
\label{fig:var}
\end{figure}

\begin{figure}[H]
\centering
\begin{subfigure}{.3\textwidth}
  \centering
  \includegraphics[width=1.0\linewidth]{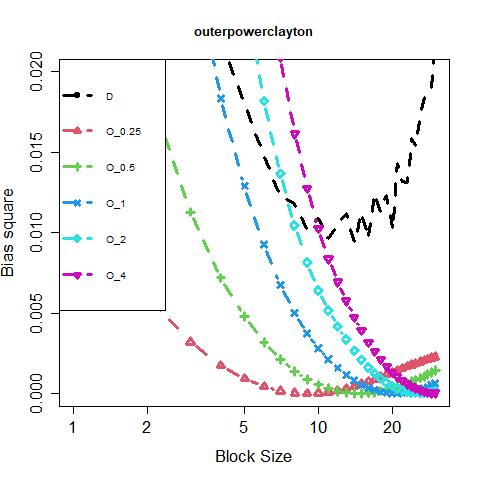}
\end{subfigure}%
\begin{subfigure}{.3\textwidth}
  \centering
  \includegraphics[width=1.0\linewidth]{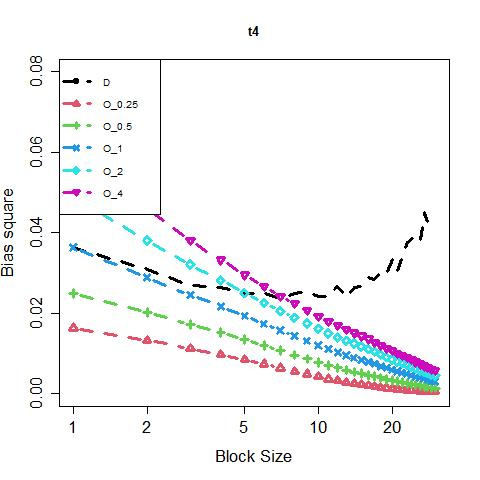}
\end{subfigure}%
\begin{subfigure}{.3\textwidth}
  \centering
  \includegraphics[width=1.0\linewidth]{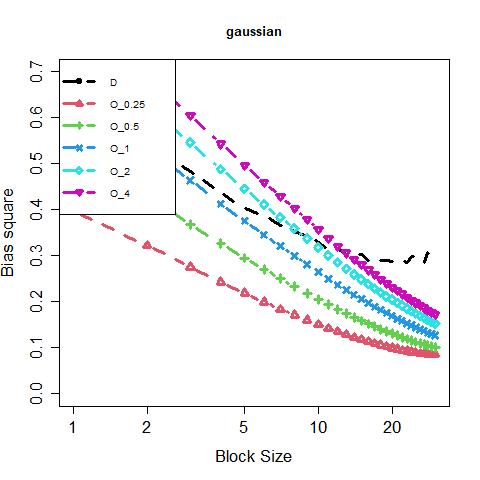}
\end{subfigure}
\caption{$B^{(sum)}$ of estimators in \Cref{tab:est} under outer-power Clayton, $t_{4}$, and Gaussian copulas}
\label{fig:bias}
\end{figure}

We examine the criteria below by setting T = 51 below and by averaging out over N = 1000 iterations:
\begin{align*}
B^{(sum)}&=\sum_{t=0}^{T-1}\bigg\{\hat{\Exp}\Big[\check{A}_{m}^{\Psi}\big(t/(T-1)\big)-A\big(t/(T-1)\big)\Big]\bigg\}^{2},\\
Var^{(sum)}&=\sum_{t=0}^{T-1}\widehat{\Var}\Big(\check{A}_{m}^{\Psi}\big(t/(T-1)\big)\Big),\\
MSE^{(sum)}&=B^{(sum)}+Var^{(sum)}.
\end{align*}
\Cref{fig:mse,fig:bias,fig:var} include $B^{(sum)}$, $Var^{(sum)}$, and $MSE^{(sum)}$ of the estimators in \Cref{tab:est}.

\begin{table}[H]
\begin{center}
\begin{tabular}{lll}
Disjoint/Overlap & Value of $c$ & Shorthand \\ \hline
Disjoint         & 1            & D         \\
Overlap          & 0.25         & O\_0.25   \\
Overlap          & 0.5          & O\_0.5    \\
Overlap          & 1            & O\_1      \\
Overlap          & 2            & O\_2      \\
Overlap          & 4            & O\_4     
\end{tabular}
\end{center}
\caption{Estimators for $A_{\infty}$ and their shorthands}
\label{tab:est}
\end{table}

\Cref{fig:mse,fig:bias,fig:var} show that having a smaller constant $c$, or equivalently, having a higher ``threshold'', increases the variance, reduces the bias, and can potentially diminish the overall MSE. Particularly, the overlapping-block estimator with $c=0.25$ corresponds to MSE curves with the smallest nadirs. Data-adaptive selections of the combination of parameter $c$ and block size $m$ will be left to future works. 

\section*{Appendix}

\begin{proof}[Proof of \Cref{prop:BiasVar}(i)]

By \cite[(B.1) and Lemma 3.4]{zou2021multiple}, \eqref{eq:extcopula},  \Cref{assump:high_order}, and \Cref{assump:AlphaMixing}(iv), for all $s>0$ and $(u,v)\in (0,1]^{2}$,
\[
S(u^{s},v^{s})=s^{1-\rho}S(u,v)\frac{C_{\infty}(u^{s},v^{s})}{C_{\infty}(u,v)}=s^{1-\rho}S(u,v)\bigg(C_{\infty}(u,v)\bigg)^{s-1}.
\]
By \eqref{eq:copula_in_pickands}, $C_{\infty}\Big(e^{-(1-t)},e^{-t}\Big)=e^{-A_{\infty}(t)}.$
Hence, for $c>0$ and $0<y<1$,
\begin{equation*}
\begin{aligned}
S\Big(y^{c(1-t)}, y^{ct}\Big)&=S\Big(e^{-c\log (y)(-(1-t))}, e^{-c\log (y)(-t)}\Big)\\
&=\big(-c\log (y)\big)^{1-\rho}S\Big(e^{-(1-t)},e^{-t}\Big)\bigg(C_{\infty}\Big(e^{-(1-t)},e^{-t}\Big)\bigg)^{-c\log (y)-1}\\
&=\big(-c\log (y)\big)^{1-\rho}S\Big(e^{-(1-t)},e^{-t}\Big)e^{(c\log (y)+1)A_{\infty}(t)}.
\end{aligned}
\end{equation*}
Hence, by the uniform convergence in \eqref{eq:secor},
\begin{equation*}\label{eq:bias}
\begin{aligned}
\Exp\Big[\hcS(t,c)-\cS(t,c)\Big]&=-\int_{0}^{1}C_{m}(y^{c(1-t)}, y^{ct})-C_{\infty}(y^{c(1-t)}, y^{ct})\diff y\\
&=-\ca(m)\int_{0}^{1}S\Big(y^{c(1-t)}, y^{ct}\Big)\diff y +o(\ca(m))\\
&=-\ca(m)S\Big(e^{-(1-t)},e^{-t}\Big)e^{A_{\infty}(t)}c^{1-\rho}\int_{0}^{1}\big(-\log (y)\big)^{1-\rho}y^{cA_{\infty}(t)}\diff y+o(\ca(m))\\
&=-\ca(m)S\Big(e^{-(1-t)},e^{-t}\Big)e^{A_{\infty}(t)}\Gamma(2-\rho)c^{1-\rho}\Big(cA_{\infty}(t)+1\Big)^{\rho-2}+o(\ca(m)),\\
\end{aligned}    
\end{equation*}
where $\Gamma(\cdot)$ is the Gamma function. By a Taylor expansion, the Dominated Convergence Theorem, and \eqref{eq:pickands_in_mado}, 
\begin{equation*}
\begin{aligned}
\left|\Exp\big[\hA_{\infty}(t)-A_{\infty}(t)\big]\right|&=\left|\frac{1}{c}\Exp\Bigg[\bigg(\frac{1}{1-\hcS(t,c)}-1\bigg)-\bigg(\frac{1}{1-\cS(t,c)}-1\bigg)\Bigg]\right|\\
&=\left|\frac{1}{c}\Bigg[\Big(1-\cS(t,c)\Big)^{-2}+o(1)\Bigg]\Exp\Big[\hcS(t,c)-\cS(t,c)\Big]\right|\\
&=\ca(m)S\Big(e^{-(1-t)},e^{-t}\Big)e^{A_{\infty}(t)}\Gamma(2-\rho)\Bigg(\frac{cA_{\infty}(t)+1}{c}\Bigg)^{\rho}+o(\ca(m)).
\end{aligned}
\end{equation*}
\end{proof}

\begin{proof}[Proof of \Cref{prop:BiasVar}(ii)]
By \Cref{assump:AlphaMixing}(i)-(iii) and a proof similar to \cite[Theorem 3.1]{Buc14},
\begin{align*}
\Cov\Big(\hC_{m}^{\circ}(u_{1},v_{1}),\hC_{m}^{\circ}(u_{2},v_{2})\Big) =& (m/n)\Big(C_{\infty}(\min(u_{1},v_{1}),\min(u_{2},v_{2}))-C_{\infty}(u_{1},v_{1})C_{\infty}(u_{2},v_{2})\Big)\\
&+o(m/n).
\end{align*}
Hence, by \eqref{eq:mado_in_copula},
\begin{equation*}
\begin{aligned}
\Var\Big(\hcS(t,c)\Big)&= \int_{0}^{1}\int_{0}^{1}\Cov\Big(\hC_{m}^{\circ}(y^{c(1-t)},y^{ct}), \hC_{m}^{\circ}(z^{c(1-t)},z^{ct}) \Big) \diff z \diff y \\
&= 
(m/n)\bigg\{2\int_{0}^{1}\int_{y}^{1} C_{\infty}\Big(y^{c(1-t)},y^{ct}\Big) \diff z \diff y -
\Big[\int_{0}^{1}C_{\infty}\Big(y^{c(1-t)},y^{ct}\Big) \diff y \Big]^{2}\bigg\}+o(m/n)\\
&=(m/n)\bigg\{2\Big[1-\cS(t,c)\Big]-\Big[1-\cS(t,c/2)\Big]-\Big[1-\cS(t,c)\Big]^{2}\bigg\}+o(m/n)\\
&=(m/n)\frac{cA_{\infty}(t)}{\big(cA_{\infty}(t)+1\big)^{2}\big(cA_{\infty}(t)+2\big)}+o(m/n).
\end{aligned}
\end{equation*}
Hence, by a Taylor expansion, the Dominated Convergence Theorem, and \eqref{eq:pickands_in_mado},
\begin{equation*}
\begin{aligned}
    \Var\big(\hA_{\infty}(t)\big)&=\frac{1}{c^{2}}\Var\Bigg[\bigg(\frac{1}{1-\hcS(t,c)}-1\bigg)-\bigg(\frac{1}{1-\cS(t,c)}-1\bigg)\Bigg]\\
&=\frac{1}{c^{2}}\Big[\Big(1-\cS(t,c)\Big)^{-2}+o(1)\Big]^{2}\Var\Big(\hcS(t,c)-\cS(t,c)\Big)\\
&=(m/n)\frac{\big(cA_{\infty}(t)+1\big)^{2}A_{\infty}(t)}{c\big(cA_{\infty}(t)+2\big)}+o(m/n).
\end{aligned}
\end{equation*}
\end{proof}

\section*{Acknowledgments}
The author would like to thank Stanislav Volgushev for fruitful discussions. 
\printbibliography

@article{bucher2021horse,
  title={A Horse Race between the Block Maxima Method and the Peak--over--Threshold Approach},
  author={B{\"u}cher, Axel and Zhou, Chen},
  journal={Statistical Science},
  volume={36},
  number={3},
  pages={360--378},
  year={2021},
  publisher={Institute of Mathematical Statistics}
}

@article{Buc14,
  title={Extreme value copula estimation based on block maxima of a multivariate stationary time series},
  author={B{\"u}cher, Axel and Segers, Johan},
  journal={Extremes},
  volume={17},
  number={3},
  pages={495--528},
  year={2014},
  publisher={Springer}
}

@article{berghaus2013minimum,
  title={Minimum distance estimators of the Pickands dependence function and related tests of multivariate extreme-value dependence},
  author={Berghaus, Betina and B{\"u}cher, Axel and Dette, Holger},
  journal={Journal de la Soci{\'e}t{\'e} Fran{\c{c}}aise de Statistique},
  volume={154},
  number={1},
  pages={116--137},
  year={2013}
}

@article{bucher2011new,
  title={New estimators of the Pickands dependence function and a test for extreme-value dependence},
  author={B{\"u}cher, Axel and Dette, Holger and Volgushev, Stanislav},
  journal={The Annals of Statistics},
  pages={1963--2006},
  year={2011},
  publisher={JSTOR}
}

@article{caperaa1997nonparametric,
  title={A nonparametric estimation procedure for bivariate extreme value copulas},
  author={Cap{\'e}ra{\`a}, Philippe and Foug{\`e}res, A-L and Genest, Christian},
  journal={Biometrika},
  volume={84},
  number={3},
  pages={567--577},
  year={1997},
  publisher={Oxford University Press}
}

@article{cormier2014using,
  title={Using B-splines for nonparametric inference on bivariate extreme-value copulas},
  author={Cormier, Eric and Genest, Christian and Ne{\v{s}}lehov{\'a}, Johanna G},
  journal={Extremes},
  volume={17},
  number={4},
  pages={633--659},
  year={2014},
  publisher={Springer}
}

@article{escobar2018local,
  title={Local robust estimation of the Pickands dependence function},
  author={Escobar-Bach, Mikael and Goegebeur, Yuri and Guillou, Armelle},
  journal={The Annals of Statistics},
  volume={46},
  number={6A},
  pages={2806--2843},
  year={2018},
  publisher={Institute of Mathematical Statistics}
}

@article{fonseca2015generalized,
  title={Generalized madogram and pairwise dependence of maxima over two regions of a random field},
  author={Fonseca, Cec{\'\i}lia and Pereira, Lu{\'\i}sa and Ferreira, Helena and Martins, Ana Paula},
  journal={Kybernetika},
  volume={51},
  number={2},
  pages={193--211},
  year={2015},
  publisher={Institute of Information Theory and Automation AS CR}
}

@article{genest2009rank,
  title={Rank-based inference for bivariate extreme-value copulas},
  author={Genest, Christian and Segers, Johan},
  journal={The Annals of Statistics},
  volume={37},
  number={5B},
  pages={2990--3022},
  year={2009},
  publisher={Institute of Mathematical Statistics}
}

@article{gudendorf2012nonparametric,
  title={Nonparametric estimation of multivariate extreme-value copulas},
  author={Gudendorf, Gordon and Segers, Johan},
  journal={Journal of Statistical Planning and Inference},
  volume={142},
  number={12},
  pages={3073--3085},
  year={2012},
  publisher={Elsevier}
}

@article{guillou2014madogram,
  title={Madogram and asymptotic independence among maxima},
  author={Guillou, Armelle and Naveau, Philippe and Schorgen, Antoine},
  journal={REVSTAT-Statistical Journal},
  volume={12},
  number={2},
  year={2014}
}

@article{hall2000distribution,
  title={Distribution and dependence-function estimation for bivariate extreme-value distributions},
  author={Hall, Peter and Tajvidi, Nader},
  journal={Bernoulli},
  pages={835--844},
  year={2000},
  publisher={JSTOR}
}

@article{marcon2017multivariate,
  title={Multivariate nonparametric estimation of the Pickands dependence function using Bernstein polynomials},
  author={Marcon, Giulia and Padoan, SA and Naveau, Philippe and Muliere, Pietro and Segers, Johan},
  journal={Journal of Statistical Planning and Inference},
  volume={183},
  pages={1--17},
  year={2017},
  publisher={Elsevier}
}

@article{naveau2009modelling,
  title={Modelling pairwise dependence of maxima in space},
  author={Naveau, Philippe and Guillou, Armelle and Cooley, Daniel and Diebolt, Jean},
  journal={Biometrika},
  volume={96},
  number={1},
  pages={1--17},
  year={2009},
  publisher={Oxford University Press}
}

@article{peng2013weighted,
  title={Weighted estimation of the dependence function for an extreme-value distribution},
  author={Peng, Liang and Qian, Linyi and Yang, Jingping},
  journal={Bernoulli},
  volume={19},
  number={2},
  pages={492--520},
  year={2013},
  publisher={Bernoulli Society for Mathematical Statistics and Probability}
}

@inproceedings {Pic81,
    AUTHOR = {Pickands, III, James},
     TITLE = {Multivariate extreme value distributions},
 BOOKTITLE = {Proceedings of the 43rd session of the {I}nternational
              {S}tatistical {I}nstitute, {V}ol.\ 2 ({B}uenos {A}ires, 1981)},
      NOTE = {With a discussion},
   JOURNAL = {Bull. Inst. Internat. Statist.},
  FJOURNAL = {Bulletin de l'Institut International de Statistique},
    VOLUME = {49},
      YEAR = {1981},
     PAGES = {859--878, 894--902},
     CODEN = {BISQA3},
   MRCLASS = {60F05 (62G30 62H99)},
  MRNUMBER = {820979},
}

@article{segers2012nonparametric,
  title={Nonparametric inference for max-stable dependence},
  author={Segers, Johan},
  journal={Statistical Science},
  volume={27},
  number={2},
  pages={193--196},
  year={2012},
  publisher={JSTOR}
}

@article{vettori2018comparison,
  title={A comparison of dependence function estimators in multivariate extremes},
  author={Vettori, Sabrina and Huser, Rapha{\"e}l and Genton, Marc G},
  journal={Statistics and Computing},
  volume={28},
  number={3},
  pages={525--538},
  year={2018},
  publisher={Springer}
}

@article{zou2021multiple,
  title={Multiple block sizes and overlapping blocks for multivariate time series extremes},
  author={Zou, Nan and Volgushev, Stanislav and B{\"u}cher, Axel},
  journal={The Annals of Statistics},
  volume={49},
  number={1},
  pages={295--320},
  year={2021},
  publisher={Institute of Mathematical Statistics}
}
\end{document}